\documentclass[manuscript]{emulateapj}
\usepackage{times}

\slugcomment{submitted for publication to ApJ Letters}

\shorttitle{Faint red upturn in A1689}
\shortauthors{Ba{\~n}ados et al.}

\begin{document}


\title{The faint end of the galaxy luminosity function in Abell 1689:
a steep red faint end upturn at $z=0.18$}


\author{Eduardo Ba{\~n}ados\altaffilmark{1}, Li-Wei Hung\altaffilmark{2},
Roberto De Propris\altaffilmark{3}, Michael J. West\altaffilmark{4}}


\altaffiltext{1}{Departamento de Astronomia y Astrofisica, Pontificia Universidad
Catolica de Chile, Santiago, Chile}
\altaffiltext{2}{Department of Astronomy, Ohio State University, Columbus, OH, USA}
\altaffiltext{3}{Cerro Tololo Inter-American Observatory, La Serena, Chile}
\altaffiltext{4}{European Southern Observatory, Santiago, Chile}


\begin{abstract}

We present a deep and wide $I$ luminosity function for galaxies in Abell 1689 ($z=0.183$)
from a mosaic of HST WFPC2 images covering $10'$ on the side. The main result of this
work is the detection of a steep upturn in the dwarf galaxy LF, with $\alpha \sim -2$. 
The dwarf to giant ratio appears to increase outwards, but this is because giant galaxies
are missing in the cluster outskirts, indicating luminosity segregation. The red sequence
LF has the same parameters, within errors, as the total LF, showing that the faint end
upturn consists of red quiescent galaxies. We speculate that the upturn is connected to
the `filling-in' of the red sequence at $z < 0.4$ and may represent the latest installment
of `downsizing' as the least massive galaxies are being quenched at the present epoch.
\end{abstract}





\section{Introduction}

The luminosity function (hereafter LF) of galaxies provides a powerful handle to
understand galaxy formation and evolution. In the usual Schechter form, the
characteristic luminosity $L^*$ may be taken as a measure of the mean luminosity
of giant galaxies, while the slope $\alpha$ measures the relative abundance 
of dwarf galaxies. The variation of these two parameters with redshift and
environment yields a measure of the growth (in terms of stellar light and mass,
although the relationship between these two quantities is not straightforward)
of luminous objects and probes the evolution of the dwarf population. As a zeroth 
order description  of galaxy properties, the LF is both an essential ingredient and 
an important test for models of galaxy formation (e.g., see \citealt{bower2010}).

The LF is measured most economically in clusters of galaxies, whose members
can be enumerated statistically or distinguished on the basis of their characteristic
colors and morphologies. One advantage of clusters is that we may consider
their populations to constitute a volume-limited sample of galaxies observed
at the same cosmic epoch and in an environment that corresponds to the densest
peaks in the dark matter distribution at each epoch. Clusters of galaxies then may
represent a snapshot of the evolving galaxy population out to very high lookback
times and their LFs allow us to reconstruct the history of galaxy formation.

The observational consensus is that giant galaxies have assembled most of their
mass by $z \sim 1.5$ \citep{depropris99,andreon06,depropris07,muzzin08} and
also have formed their stellar populations rapidly and at $z > 2.5$ \citep{blakeslee03,
mei06a,mei06b,mei09}. The behavior of dwarf galaxies is however not yet as well
understood. In $z > 0.4$ clusters from the ESO Distant Cluster Survey (hereafter EDisCS),
\cite{delucia07} find that the red sequence is weaker at lower luminosities, indicating
a relative deficit of quiescent dwarf galaxies compared to present-day clusters. This 
result is still somewhat controversial, with some further studies confirming or even
strengthening the observed deficit (e.g., \citealt{stott07,gilbank08,hansen09})
but others finding no evidence for evolution of the faint end of the red sequence
(e.g., \citealt{andreon08}, Crawford, Bershady \& Hoessel 2009) and arguing that
the apparent lack of faint red galaxies may be due to selection effects and/or cluster
to cluster variations (cf., the reanalysis of the original \citealt{stott07} data by
Capozzi, Collins \& Stott 2010). 

The deficit of faint red galaxies in clusters may represent a cluster version of `downsizing': 
dwarfs may either reside in the cluster blue cloud and migrate to the red sequence once their 
star formation is quenched, or may be accreted from the general field. Locally, there is evidence 
that at least some of the fainter dwarfs in the Virgo cluster were forming stars until recently
(Jerjen, Kalnajs \& Binggeli 2000; Conselice, Gallagher \& Wyse 2001; \citealt{janz09}), while
in the  Coma cluster \cite{smith09} find that dwarf galaxies span a wide range of ages: however, 
dwarfs  in the cluster core are generally as old as the giants, while a younger population is 
present in the outskirts, consistent with recent infall from the surrounding field. 

Also of importance is the slope of the {\it total} LF. Although this is more difficult to
measure, it is this quantity that would allow us to answer the question whether the fainter 
dwarf galaxies are already present in clusters but lie on the blue sequence at the EDisCS
epoch, or they have been recently accreted. Determining the total LF slope at high redshift 
is a more complex proposition, because of the lower contrast against the foreground and 
background contamination compared to the small color range sampled by the red sequence. This 
requires deep and wide field observations of several clusters, with homogeneous imaging of 
`blank' fields to assess the number of contaminating objects statistically, and/or numerous 
bandpasses to perform photometric redshift analysis.

We have begun a project to determine the evolution of the LF in $z > 0.2$ clusters using 
archival data from the Hubble Space Telescope (HST): most of the original observations we
use were taken to study galaxy lensing and are therefore deep and cover enough area
to sample the faint end of the LF even at large distances from the cluster centres. With these
data we can exploit HST's superior photometric performance and stability, small point
spread function (especially important for dwarf galaxies), high spatial resolution, and 
the availability of numerous deep `blank' fields (e.g., COSMOS, EGS, GOODS) to provide
a homogeneous set of data (taken under the same conditions) for statistical subtraction
of foreground and background galaxies lying along the cluster line of sight.

In \cite{pracy04} we applied this method to a wide $R$ band mosaic of WFPC2 fields in
Abell 2218 and derived a LF down to $M_R \sim -12 + 5 \log h$, showing that the LF
slope appears to steepen outwards and that the faintest dwarfs avoid the cluster centre.
\cite{harsono07,harsono09} have derived a deep composite LF in six bands for five
clusters at $<z>=0.25$ and find that the population of dwarf galaxies down to $M_z=
-14 + 5\log h$ was already present, fully assembled and lying on the red sequence 
at $z \approx 0.3$, but find no faint end upturn. Here we present a study of the faint 
end of the $I$ band (F814W) LF in Abell 1689 ($z=0.183$) over a $10'$ field imaged with 
HST WFPC2. The next section describes the data and photometry, while we present our results 
and discussion in the following sections. We adopt the WMAP7 cosmological parameters: $\Omega_M
=0.27$, $\Omega_{\Lambda}=0.73$ and H$_0=71$ km s$^{-1}$ Mpc$^{-1}$.

\section{Data Analysis}

The data used in this paper consist of a $4 \times 4$ WFPC2 mosaic of Abell 1689
covering $\sim 10' \times 10'$ on the sky with exposure times of 1800s in the $V$
(F606W) band and 2300s in the $I$ (F814W) band. The images were retrieved as fully
processed and drizzled files from the HST Legacy Archive (PID: 5993; PI: Kaiser).

In order to determine the LF of cluster galaxies we need to estimate the contribution
to the total galaxy counts in the cluster line of sight from galaxies in the field (in the
foreground or background). We use the $I$ band counts in the COSMOS field \citep{leauthaud07}. 
These counts have similar photometric depth to our data (the exposure times are similar, but the ACS
is about a factor of 2 more efficient than WFPC2), cover a large area (1.64 deg$^2$; therefore
minimizing the effects of cosmic variance) and are taken in a closely related filter. We therefore expect that we can use these counts to decontaminate our dataset statistically and recover the 
LF of cluster members. 

For consistency, we analyze our data in the same manner as \cite{leauthaud07}: we run the 
Sextractor \citep{bertin96} package twice and with the same parameters as used for the COSMOS 
field. A first pass with coarse search parameters is used to detect the bright galaxies without deblending them, while a second pass with finer search parameters is used for the faint galaxies. 
All detections were visually inspected to remove spurious sources, artifacts and especially
arclets. All photometry was calibrated to the AB system using published zeropoints. By this
approach we are able to use the COSMOS counts for our background removal. In addition, we 
also measured two aperture magnitudes in $V$ and $I$ (in an aperture equivalent to 5 $h^{-1}$ 
kpc) in order to determine the galaxy colors, identify the red sequence, and use this to estimate 
the red sequence luminosity function. We also measure the ellipticity and position angle of 
galaxies: these are used for a companion paper on the alignment effect (Hung et al. 2010, 
in preparation). 

Star-galaxy separation is carried out using the $\mu_{max}$ vs. $I$ diagram shown in
Figure 1 \citep{leauthaud07}, where $\mu_{max}$ is the central surface brightness of
each object. Stars will define a tight sequence in this plot, while galaxies will occupy 
a cloud of points at lower $\mu_{max}$ for a given total luminosity. The adopted discriminant
between stars and galaxies is shown in Figure 1. As a check on our method, we determined 
galaxy counts in the $I$ band for the two Hubble Deep Fields \citep{williams96,casertano00} 
and verified that we obtain good agreement with the COSMOS counts reported by 
\cite{leauthaud07}.

\begin{figure}
\epsscale{.80}
\plotone{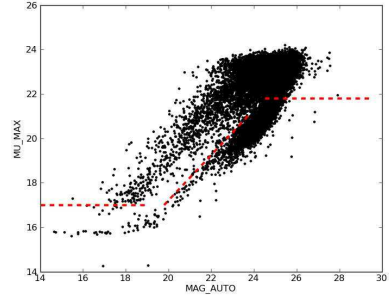}
\caption{
Star-galaxy separation indices: we plot $\mu_{max}$ (the maximum surface
brightness for each object, usually at its centroid) vs. total magnitude $I$
for all confirmed detections in the Abell 1689 field. In this plot, stars
define a tight sequence while galaxies will occupy a cloud of points at
lower $\mu_{max}$ at each $I$. The thick red dashed line shows the adopted
discrimination between stars and galaxies. Objects 'below' and to the right of this
line are classified as stars. The separation is reliable to $I \sim 24$.
}
\vspace{1cm}
\end{figure}

\section{The Luminosity Functions of Galaxies in Abell 1689}

In Figure 2(a) we show the LF of galaxies in Abell 1689 for the entire area covered by the
WFPC2 observations. We subtracted the scaled fore/back-ground counts derived from
the COSMOS field, assuming Poissonian errors for the galaxy counts and including terms
due to clustering errors (as per \citealt{huang97,driver03}). All errors are added in quadrature.
It is clear that the data are not a good fit to a standard Schechter function: the LF appears to
flatten at intermediate magnitudes and presents a steep rise at faint luminosities. This is 
similar to the local deep composite LFs for Sloan and RASS clusters observed by \cite{popesso06}
and to the original claims for a steep faint end upturn of the LF in clusters of galaxies
\citep{driver94,depropris95}.

\begin{figure}
\epsscale{.80}
\plotone{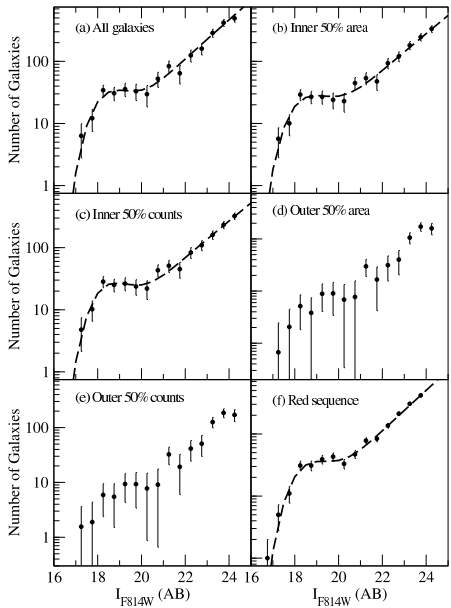}
\caption{Luminosity functions and best fits for Abell 1689. Panel (a) is the total LF;
panel (b) shows the LF for galaxies in the inner 50\% of the area covered; panel (c)
is for galaxies in the area containing 50\% of the counts; panels (d) and (e) are the
same as panels (b) and (c) but for the outer 50\% regions; panel (f) is the total LF
from the red sequence.}
\end{figure}

Following \cite{popesso06}, we fit our data with a combination of a Schechter function and a 
power law:

\begin{eqnarray}
\Phi(M)dM=\Phi^* 10^{0.4(M^*-M)(\alpha+1)} \exp(-10^{0.4(M^*-M)}) \times \nonumber \\
(1+10^{0.4(M_t-M)\beta})dM \nonumber
\end{eqnarray}

where $\Phi^*$, $M^*$ and $\alpha$ are the usual Schechter function parameters, $M_t$ is
the transition magnitude between the Schechter function and power-law behavior and $\beta$
is the index of the power-law at the faint end. The best fitting values are reported in 
Table 1. The errors on each parameter are derived by holding each of the other parameters
fixed. The table columns are: the region over which the LF is determined, the four 
parameters that determine its shape ($M^*$, $\alpha$, $M_t$, $\beta$) and their errors
and the dwarf to giant ratio (as defined below). This appears to indicate the presence 
of a steep upturn in Abell 1689, as earlier found by \cite{wilson97} using ground-based 
data in the $V$ band.

The deficit (or otherwise) of faint red galaxies is measured using the dwarf to giant 
ratio, i.e., the ratio of the numbers of galaxies within specified luminosity intervals.
In our case the most appropriate intervals may be defined from the transition magnitude 
between the Schechter and power-law behavior in the LF. We take 'giants' to be galaxies 
with $M_I > -19.5$ and 'dwarfs' to be galaxies with $M_I < -19.5$ and tabulate the derived 
ratio in Table 1. This is different than the definition used by \cite{delucia07} and others, 
but allows us to measure the importance of the faint end upturn as a function of cluster-centric
radius and may be interpreted as an estimate of the relative strength of the separate dwarf
and giant populations, which obey different LFs (e.g., Binggeli, Sandage \& Tammann 1988; 
\citealt{lu09}). 

In our previous work \cite{pracy04} found a trend for the LF slope to steepen in the
outskirts of Abell 2218 (however, there was no prominent upturn as in Abell 1689 and
the LF was well fitted by a single Schechter function), with the faintest dwarf galaxies
preferentially avoiding the central region. In Figures 2(b) and 2(c) we plot the LFs for 
galaxies within the region containing 50\% of the area surveyed and 50\% of the galaxy 
counts, respectively. These correspond to a cluster-centric radius of 590 and 630 kpc, 
respectively, or around 20\% of the cluster virial radius ($r_{200}$). The LF parameters 
and dwarf to giant ratios for these LFs are tabulated in Table 1. These LFs are consistent 
with each other and with the presence of a steep upturn.

\begin{deluxetable*}{cccccc}
\tablecaption{Derived LF parameters}
\tablewidth{0pt}
\tablehead{
\colhead{Region} & \colhead{$M^*$} & \colhead{$\alpha$} &
\colhead{$M_t$} & \colhead{$\beta$} & \colhead{D/G ratio}
}
\startdata
Entire Field & $19.01 \pm 0.26$ & $0.33 \pm 0.49$ & $M_t=19.92  \pm 0.27$ & 
$-2.09 \pm 0.44$ & $8.0 \pm 0.7$ \\
Inner 50\% Area & $18.88 \pm 0.34$ & $0.17 \pm 0.74$ & $20.03 \pm 0.23$ & 
$-1.90 \pm 0.66$ & $6.6 \pm 0.6$ \\
Inner 50\% Counts & $18.99 \pm 023$ & $0.35 \pm 0.44$ & $19.97 \pm 0.21$ & 
$-2.07 \pm 0.41$ & $6.0 \pm 0.6$ \\
Outer 50\% Area & \nodata & \nodata & \nodata & $-1.5 \pm 0.1$ & $13.9 \pm 2.6$ \\
Outer 50\% Counts & \nodata & \nodata & \nodata & $-1.5 \pm 0.1$ & $14.1 \pm 2.5$ \\
Red sequence & $19.01 \pm 0.29$ & $0.17 \pm 0.43$ & $20.03 \pm 0.21$ & $-1.97 \pm
0.25$ & $8.2 \pm 0.7$ \\
\enddata
\end{deluxetable*}

In panels 2(d) and 2(e) we show the LFs for the outer 50\% area and that containing the 
remaining 50\% of the counts respectively. The areas covered go out to the edge of the 
observed mosaic (about 1 Mpc for the chosen cosmological parameters or 32\% of the virial
radius). The brighter galaxies appear to be substantially deficient in these regions and 
we are only able to fit a power law to galaxies fainter than $I=21$ where we should probe 
the $\beta$ parameter of the LF in equation 1. The best fit for both these LF has $\beta 
\sim -1.5 \pm 0.1$. The derived dwarf to giant ratio is higher (by about a factor of 2) than 
in the two inner regions. However, the LF slope is shallower than in the central 600 kpc. 
The increase in the dwarf to giant ratio is therefore due to a relative lack of giants rather 
than to an increased dwarf contribution, consistent with luminosity segregation.

\section{Discussion}

We have found a steep upturn in the $I$ band LF of Abell 1689 and shown that this upturn
extends throughout the inner 600 kpc of this cluster. The LF is steep in this region
but becomes shallower in the cluster outskirts. It is now clear that an upward inflection 
of the LF is common for clusters at $z < 0.2$ \citep{popesso06}.

What is the nature of the faint upturn population ? In the clusters surveyed by 
\cite{popesso06} the similarity of LF parameters across the SDSS bands suggests 
that these galaxies are mostly red. Similarly, the red sequence LF in the lowest
redshift bin ($z=0.20$, their Fig.~16) in \cite{lu09} shows an upturn (although of
course their data reach to brighter magnitudes than ours). Abell 1689 hosts a 
prominent red sequence population (Figure 3). The previous work by \cite{wilson97} 
also finds a steep upturn for this cluster {\it in the V band} with similar parameters 
to those we derive for the $I$ band. This suggests that the upturn in Abell 1689 also
consists of galaxies on the red sequence. 

\begin{figure}
\epsscale{0.8}
\plotone{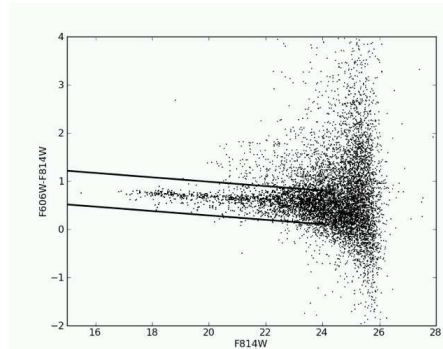}
\caption{Color-magnitude relation for Abell 1689. Selection lines show the range of
colors we adopt for cluster membership to derive the red sequence LF in Fig,~2(f).}
\end{figure}

Unfortunately, the COSMOS field has not been imaged with HST in the $V$ (F606W) band, 
although it includes extensive ground based observations. We cannot therefore carry out 
an appropriate background subtraction in the $V$ band or on the $V-I$ color-magnitude
relation. Nevertheless, we can check whether the LF on the red sequence is at least 
consistent with a steep upturn. We fit a straight line to the red sequence and assume 
that all galaxies within $\pm 0.3$ of this `ridge-line' are cluster members. The selection 
region is shown in Figure 3. Figure 2(f) shows the red sequence LF and best fit; the 
parameters are tabulated in Table 1. This also shows a steep upturn, with slope and dwarf 
to giant ratio consistent with the LFs for the entire cluster areas and those within the 
two central regions. Although this is not corrected for background contamination (which may 
produce an excessively steep LF), it suggests, together with the concordance between the 
$V$ band LF from \cite{wilson97} and ours in the $I$ band, that the faint upturn population
consists largely of red galaxies. 

Additionally, although the dwarf to giant ratio appears to increase outwards, this is due
to a reduced contribution from the giant population at large cluster-centric radii rather
than from a steep dwarf galaxy LF from newly infalling objects (see the steep LF derived
for the general field by \citealt{christlein09}). This suggests that Abell 1689 has experienced
luminosity segregation, as observed elsewhere \citep{andreon02,mercurio03}. The faint end of
the LF appears to steepen inward, arguing that the red upturn population may already have 
been present in the cluster (although it might have resided in the blue cloud at earlier
epochs), rather than having been accreted from the field. A caveat to this interpretation 
is the finding by \cite{pracy04} that the less luminous galaxies tend to avoid the cluster 
core, while \cite{harsono07,harsono09} and \cite{riley09} also find relatively flat LFs (
$\alpha \sim -1.3$) in the inner $2'$ of $z \sim 0.3$ clusters. Of course it is possible 
that the cluster core is particularly hostile to dwarf galaxies, while these may be abundant
immediately outside of the giant-dominated region, and the results may not be inconsistent 
with our findings.

If we accept the evidence for a deficit of faint red galaxies at $z > 0.4$ from the
EDisCS sample and other studies (\citealt{delucia07,gilbank08,hansen09}, modulo the
critiques of \cite{andreon08,crawford09}), then a large population of faint dwarfs
must have arrived on the red sequence between $z \sim 0.4$ and at least $z \approx 0.2$
(Abell 963, the $z=0.20$ LF in \citealt{lu09}'s Fig.~16). Our data suggest that these 
dwarf galaxies were already part of the cluster environment and therefore must have had 
their star formation quenched in the $0.2 < z < 0.4$ interval.

\cite{lu09} conclude that most dwarfs have reached the red sequence at $z < 0.2$ but find 
little evolution at $0.2 < z < 0.4$  The best comparison is provided by their $z=0.2$ LF in 
their Fig.~6. \cite{lu09} find a `dip' at moderate luminosity followed by an upturn. 
This resembles closely our total LF for Abell 1689 as well as our `best-guess' red LF in 
Fig.~2(f). According to \cite{lu09} the dwarf to giant ratio should increase at lower 
redshifts, although this is based on a comparison with different data. Abell 1689 itself
may not be the best comparison, as it may be more highly evolved (as indicated by the
observation of luminosity segregation).

The behavior of the upturn at higher redshifts cannot be determined from \cite{lu09},
whose data only reach to the LF inflection at $z > 0.3$. However, if the faint red dwarfs have
reached the upturn by $z=0.18$ and were already present in the cluster (based on the increasing
slope $\beta$ in the inner regions), the blue band LF in moderate redshift clusters should be
quite steep. Our deep data at $z \sim 0.2$ -- $0.3$ \citep{harsono07,harsono09} do not show
an upturn, but cluster cores may be hostile to dwarf galaxies.

It is tempting to speculate that the rapid onset of the red sequence and the faint end 
upturn are connected and that we may be witnessing the epoch of migration (from the blue 
cloud to the red sequence) of the faintest galaxies, progressing steadily to lower
luminosities as we come closer to the present epoch, and eventually producing a red
sequence LF resembling the original steep LF expected from CDM models. However, this 
is ultimately a question that can only be answered by more data: the Multi-Cycle treasury
program to carry out multi-wavelength observations of galaxy clusters, other archival data
and targeted programs to study the dwarf galaxy evolution in clusters, will provide further
clues to this issue.

\acknowledgments

Based on observations made with the NASA/ESA Hubble Space Telescope, and obtained from the Hubble Legacy Archive, which is a collaboration between the Space Telescope Science Institute (STScI/NASA), the Space Telescope European Coordinating Facility (ST-ECF/ESA) and the Canadian Astronomy Data Centre (CADC/NRC/CSA).


{\it Facilities:}  \facility{HST (WFPC2)}

\clearpage

\end{document}